\documentclass[english]{article}
\usepackage[T1]{fontenc}
\usepackage[latin9]{inputenc}
\usepackage{color}
\usepackage{amssymb}
\usepackage{babel}
\begin{document}

\title{\textbf{On quasi-normal modes, area quantization and Bohr correspondence
principle}}

\maketitle
\begin{center}
Dipartimento di Fisica, Scuola Superiore di Studi Universitari e Ricerca
\textquotedbl{}Santa Rita\textquotedbl{}, via San Nicola snc, 81049,
San Pietro Infine (CE) Italy; Austro-Ukrainian Institute for Science
and Technology, Institut for Theoretish Wiedner Hauptstrasse 8-10/136,
A-1040, Wien, Austria; International Institute for Applicable Mathematics
\& Information Sciences (IIAMIS),  Hyderabad (India) \& Udine (Italy) 
\par\end{center}

\begin{center}
\textit{E-mail address:} \textcolor{blue}{cordac.galilei@gmail.com} 
\par\end{center}
\begin{abstract}
In Int. Journ. Mod. Phys. D 14, 181 (2005), the author Khriplovich
verbatim claims that ``the correspondence principle does not dictate
any relation between the asymptotics of quasinormal modes and the
spectrum of quantized black holes'' and that ``this belief is in
conflict with simple physical arguments''. In this paper we analyze
Khriplovich's criticisms and realize that they work only for the original
proposal by Hod, while they do not work for the improvements suggested
by Maggiore and recently finalized by the author and collaborators
through a connection between Hawking radiation and black hole (BH)
quasi-normal modes (QNMs). This is a model of quantum BH somewhat
similar to the historical semi-classical model of the structure of
a hydrogen atom introduced by Bohr in 1913. Thus, QNMs can be really
interpreted as BH quantum levels (the ``electrons'' of the ``Bohr-like
BH model''). 

Our results have also important implications on the BH information
puzzle.
\end{abstract}

\section{Introduction}

BH QNMs are frequencies of radial spin $j=0,1,2\:$ for scalar, vector
and gravitational perturbation respectively, obeying a time independent
Schröedinger-like equation \cite{key-1,key-2}. Such BH modes of energy
dissipation have a frequency which is allowed to be complex \cite{key-1,key-2}.
In a remarkable paper \cite{key-3}, York proposed the intriguing
idea to model the quantum BH in terms of BH QNMs. More recently, by
using Bohr's Correspondence Principle, Hod proposed that QNMs should
release information about the area quantization as QNMs are associated
to absorption of particles \cite{key-4,key-5}. Hod's work was improved
by Maggiore \cite{key-6} who solved some important problems. On the
other hand, as QNMs are countable frequencies, ideas on the continuous
character of Hawking radiation did not agree with attempts to interpret
QNMs in terms of emitted quanta, preventing to associate QNMs modes
to Hawking radiation \cite{key-1}. Recently, Zhang, Cai, Zhan and
You \cite{key-7,key-8,key-9,key-10} and the author and collaborators
\cite{key-11,key-12,key-13,key-14} observed that the non-thermal
spectrum by Parikh and Wilczek \cite{key-21,key-22} also implies
the countable character of subsequent emissions of Hawking quanta.
This issue enables a natural correspondence between QNMs and Hawking
radiation, permitting to interpret QNMs also in terms of emitted energies
\cite{key-11,key-12,key-13,key-14}. The model that has been developed
in \cite{key-11,key-12,key-13,key-14} is a BH model somewhat similar
to the historical semi-classical model of the structure of a hydrogen
atom introduced by Bohr in 1913 \cite{key-17,key-18}. In fact, in
the Bohr-like model for BHs in \cite{key-11,key-12,key-13,key-14}
QNMs represent the \textquotedbl{}electron\textquotedbl{} which jumps
from a level to another one and the absolute values of the QNMs frequencies
represent the energy \textquotedbl{}shells\textquotedbl{}. The emission
of Hawking quanta and the absorptions of particles represent, in turn,
the jumps among the various quantum levels. The Bohr-like model for
BHs has also important implications on the BH information paradox
\cite{key-25}.

\section{Hod's original proposal}

For Schwarzschild BH and in strictly thermal approximation, QNMs are
usually labelled as $\omega_{nl},$ where $n$ and $l$ are the ``overtone''
and the angular momentum quantum numbers \cite{key-1,key-2,key-4,key-5,key-6}.
For each $l$$\geq2$ for BH perturbations, we have a countable infinity
of QNMs, labelled by $n$ ($n=1,2,...$) \cite{key-1,key-2,key-4,key-5,key-6}.
Working with $G=c=k_{B}=\hbar=\frac{1}{4\pi\epsilon_{0}}=1$ (Planck
units), for large $n$ BH QNMs become independent of $l$ having the
structure \cite{key-1,key-2,key-4,key-5,key-6} 
\begin{equation}
\begin{array}{c}
\omega_{n}=\ln3\times T_{H}+2\pi i(n+\frac{1}{2})\times T_{H}+\mathcal{O}(n^{-\frac{1}{2}})=\\
\\
=\frac{\ln3}{8\pi M}+\frac{2\pi i}{8\pi M}(n+\frac{1}{2})+\mathcal{O}(n^{-\frac{1}{2}}).
\end{array}\label{eq: quasinormal modes}
\end{equation}
Although the behavior (\ref{eq: quasinormal modes}) is independent
of the orbital angular momentum $l$ the detailed values are irregular
and $l$-dependent for non very large imaginary part of $\omega_{n}$
\cite{key-1}. The number of quasi-normal frequencies is infinite
and the real part in (\ref{eq: quasinormal modes}) equals \cite{key-1}
\begin{equation}
\frac{\ln3}{4\pi}\approx0.087424.\label{eq: costante}
\end{equation}
Hod was the first author to note that the constant (\ref{eq: costante})
can be written in terms of $\ln3$ and proposed an intriguing interpretation
in order to explain this fact \cite{key-4,key-5}, as we will see
below. On the other hand, the imaginary part of (\ref{eq: quasinormal modes})
can be easily understood \cite{key-2}. The quasi-normal frequencies
determine the position of poles of a Green's function on the given
background and the Euclidean BH solution converges to a thermal circle
at infinity with the inverse temperature $\beta_{H}=\frac{1}{T_{H}}$
\cite{key-2}. Thus, the spacing of the poles in eq. (\ref{eq: quasinormal modes})
coincides with the spacing $2\pi iT_{H}$ expected for a thermal Green's
function \cite{key-2}.

In a famous paper, Bekenstein \cite{key-15} showed that the area
quantum of the Schwarzschild BH is $\triangle A=8\pi$ (we recall
that the \emph{Planck distance} $l_{p}=1.616\times10^{-33}\mbox{ }cm$
is equal to one in Planck units). Using properties of the spectrum
of Schwarzschild BH QNMs a different numerical coefficient has been
found by Hod in \cite{key-4,key-5}. Hod's analysis started by observing
that, as for the Schwarzschild BH the \emph{horizon area} $A$ is
related to the mass through the relation $A=16\pi M^{2},$ a variation
$\triangle M$ in the mass generates a variation

\begin{equation}
\triangle A=32\pi M\triangle M\label{eq: variazione area}
\end{equation}
in the area. By considering a transition from an unexcited BH to a
BH with very large $n$, Hod \cite{key-4,key-5} assumed \emph{Bohr's
correspondence principle }(which states that transition frequencies
at large quantum numbers should equal classical oscillation frequencies)
\cite{key-16,key-17,key-18} to be valid for large $n$ and enabled
a semi-classical description even in absence of a complete theory
of quantum gravity. In his approach, Hod \cite{key-4,key-5} assumed
that the relevant frequencies were the real part of the frequencies
(\ref{eq: quasinormal modes}). Hence, the minimum quantum which can
be absorbed in the transition is \cite{key-4,key-5}

\begin{equation}
\triangle M=\omega=\frac{\ln3}{8\pi M}.\label{eq: Hod}
\end{equation}
 This gives $\triangle A=4\ln3.$ This results was different from
the original result of Bekenstein \cite{key-15} while the presence
of the numerical factor $4\ln3$ stimulated possible connections with
loop quantum gravity \cite{key-19}.

\section{Criticisms by Khriplovich}

Hod's approach has been criticized by Khriplovich \cite{key-20},
who claims that properties of ringing frequencies cannot be related
directly to Bohr correspondence principle. Here we show that this
criticism works only for the original proposal by Hod \cite{key-4,key-5},
while it does not work for the improvements suggested by Maggiore
\cite{key-6} and recently finalized by the author and collaborators
\cite{key-11,key-12,key-13,key-14} through a connection between Hawking
radiation and BH QNMs (Bohr-like model for BHs). Let us see this issue
in detail. The criticisms by Khriplovich \cite{key-20} are essentially
the following:
\begin{enumerate}
\item The exact meaning of Bohr correspondence principle is the following.
In quantized systems, the energy jump $\triangle E$ between two neighbouring
levels with large quantum numbers i. e. between levels with $n$ and
$n+1,$ being $n\gg1$, is related to the classical frequency $\omega$
of the system by the formula \cite{key-20} 
\begin{equation}
\triangle E=\omega.\label{eq: salto}
\end{equation}
Thus, in a semi-classical approximation with $n\gg1$, the frequencies
which corresponds to transitions between energy levels with $\triangle n\ll n$
are integer multiples of the classical frequency $\omega$. Khriplovich
concludes by claiming that contrary to the assumption by Hod \cite{key-4,key-5},
in the discussed problem of a BH, large quantum numbers $n$ of Bohr
correspondence principle are unrelated to the asymptotics (\ref{eq: Hod})
of QNMs, but are quantum numbers of the BH itself. 
\item The real part QNMs does not differ appreciably from its asymptotic
value (\ref{eq: Hod}) in the whole numerically investigated range
of $n$, starting from $n\sim1.$ Meanwhile, the imaginary part grows
as $n+\frac{1}{2}$, and together with it the spectral width of QNMs
(in terms of common frequencies) also increases linearly with $n.$
In this situation, the idea that the resolution of a QNM becomes better
and better with the growth of $n,$ and that in the limit $n\rightarrow\infty$
this mode resolves an elementary edge (or site) of a quantized surface,
is not reasonable. 
\end{enumerate}

\section{Improvements by Maggiore}

\noindent Maggiore \cite{key-6} showed that the spectrum of BH QNMs
can be analysed in terms of superposition of damped oscillations,
of the form \cite{key-6} 
\begin{equation}
\exp(-i\omega_{I}t)[a\sin\omega_{R}t+b\cos\omega_{R}t]\label{eq: damped oscillations}
\end{equation}

\noindent with a spectrum of complex frequencies $\omega=\omega_{R}+i\omega_{I}.$
A damped harmonic oscillator $\mu(t)$ is governed by the equation
\cite{key-6} 
\begin{equation}
\ddot{\mu}+K\dot{\mu}+\omega_{0}^{2}\mu=F(t),\label{eq: oscillatore}
\end{equation}

\noindent where $K$ is the damping constant, $\omega_{0}$ the proper
frequency of the harmonic oscillator, and $F(t)$ an external force
per unit mass. If $F(t)\sim\delta(t),$ i.e. considering the response
to a Dirac delta function, the result for $\mu(t)$ is a superposition
of a term oscillating as $\exp(i\omega t)$ and of a term oscillating
as $\exp(-i\omega t)$, see \cite{key-6} for details. Then, the behavior
(\ref{eq: damped oscillations}) is reproduced by a damped harmonic
oscillator, through the identifications \cite{key-6} 

\noindent 
\begin{equation}
\begin{array}{ccc}
\frac{K}{2}=\omega_{I}, &  & \sqrt{\omega_{0}^{2}-\frac{K}{4}^{2}}=\omega_{R},\end{array}\label{eq: identificazioni}
\end{equation}

\noindent which gives 
\begin{equation}
\omega_{0}=|\omega|=\sqrt{\omega_{R}^{2}+\omega_{I}^{2}}.\label{eq: omega 0}
\end{equation}

\noindent In \cite{key-6} it has been emphasized that the identification
$\omega_{0}=\omega_{R}$ is correct only in the approximation $\frac{K}{2}\ll\omega_{0},$
i.e. only for very long-lived modes. For a lot of BH QNMs, for example
for highly excited modes, the opposite limit can be correct. Maggiore
\cite{key-6} used this observation to re-examine some aspects of
BH quantum physics that were discussed in previous literature assuming
that the relevant frequencies were $(\omega_{R})_{n}$ rather than
$(\omega_{0})_{n}$. One can indeed easily check that criticisms in
point 2 above have been well addressed by the observation by Maggiore
\cite{key-6}, who suggested that one must take 
\begin{equation}
\triangle M=\omega=(\omega_{0})_{n}-(\omega_{0})_{n-1},\label{eq: Maggiore}
\end{equation}
where $(\omega_{0})_{n}\equiv|\omega_{n}|,$ instead of the value
(\ref{eq: Hod}) proposed in \cite{key-4,key-5}. In fact, the imaginary
part becomes dominant for large $n$ and, in turn, the idea that the
resolution of a QNM becomes better and better with the growth of $n,$
and that in the limit $n\rightarrow\infty$ this mode resolves an
elementary edge (or site) of a quantized surface works. In that way,
the minimum quantum which can be absorbed in the transition becomes
\cite{key-6} 
\begin{equation}
\triangle M=\omega=\frac{1}{4M},\label{eq: quanto maggiore}
\end{equation}
which permits to find the original Bekensteins result $\triangle A=8\pi$. 

In next Section we will discuss a semi-classical model of quantum
BH that will permit to well address point 1 by Khriplovich

\section{The Bohr-like model for black holes}

\noindent Let us return on the connection between BH QNMs and Hawking
radiation. Working in strictly thermal approximation, one writes down
the probability of emission of Hawking quanta as \cite{key-21,key-22,key-23}
\begin{equation}
\Gamma\sim\exp(-\frac{\omega}{T_{H}}),\label{eq: hawking probability}
\end{equation}

\noindent being $T_{H}\equiv\frac{1}{8\pi M}$ the Hawking temperature
and $\omega$ the energy-frequency of the emitted radiation respectively.

\noindent The important correction by Parikh and Wilczek, due to the
BH back reaction yields \cite{key-21,key-22}

\noindent 
\begin{equation}
\Gamma\sim\exp[-\frac{\omega}{T_{H}}(1-\frac{\omega}{2M})].\label{eq: Parikh Correction}
\end{equation}

\noindent This result takes into account the BH varying geometry and
adds the term $\frac{\omega}{2M}$ like correction \cite{key-21,key-22}.
We have improved the Parikh and Wilczek framework by showing that
the probability of emission (\ref{eq: Parikh Correction}) is indeed
associated to the two \emph{non} strictly thermal distributions \cite{key-24}
\begin{equation}
<n>_{boson}=\frac{1}{\exp\left[4\pi\left(2M-\omega\right)\omega\right]-1},\;\;<n>_{fermion}=\frac{1}{\exp\left[4\pi\left(2M-\omega\right)\omega\right]+1},\label{eq: final distributions}
\end{equation}
for bosons and fermions respectively. Equations (\ref{eq: final distributions})
show an important deviation from the standard Bose-Einstein and Fermi-Dirac
distributions which are \cite{key-24} 
\begin{equation}
<n>_{boson}=\frac{1}{\exp\left(8\pi M\omega\right)-1}\;\;<n>_{fermion}=\frac{1}{\exp\left(8\pi M\omega\right)+1}.\label{eq: reqular distributions}
\end{equation}
In fact, the probability of emission of Hawking quanta found by Parikh
and Wilczek, i.e. eq. (\ref{eq: Parikh Correction}), shows that the
BH does NOT emit like a perfect black body, i.e. it has not a strictly
thermal behavior. On the other hand, the temperature in Bose-Einstein
and Fermi-Dirac distributions is a perfect black body temperature.
Thus, when we have deviations from the strictly thermal behavior,
i.e. from the perfect black body, one expects also deviations from
Bose-Einstein and Fermi-Dirac distributions. This is the reason of
the difference between eqs. (\ref{eq: final distributions}) and (\ref{eq: reqular distributions}).

\noindent It is well known that in various fields of physics and astrophysics
the deviation of the spectrum of an emitting body from the strict
black body spectrum is taken into account by introducing an \emph{effective
temperature,} which represents the temperature of a black body emitting
the same total amount of radiation. The effective temperature, which
is a frequency dependent quantity, can be introduced in BH physics
too \cite{key-11,key-12,key-13,key-14,key-24} as

\noindent 
\begin{equation}
T_{E}(\omega)\equiv\frac{2M}{2M-\omega}T_{H}=\frac{1}{4\pi(2M-\omega)}.\label{eq: Corda Temperature}
\end{equation}

\noindent Defining $\beta_{E}(\omega)\equiv\frac{1}{T_{E}(\omega)},$
one rewrites eq. (\ref{eq: Parikh Correction}) in Boltzmann-like
form as \cite{key-11,key-12,key-13,key-14,key-24}

\noindent 
\begin{equation}
\Gamma\sim\exp[-\beta_{E}(\omega)\omega]=\exp(-\frac{\omega}{T_{E}(\omega)}),\label{eq: Corda Probability}
\end{equation}

\noindent where one introduces the \emph{effective Boltzmann factor}
$\exp[-\beta_{E}(\omega)\omega]$ appropriate for a BH having an inverse
effective temperature $T_{E}(\omega)$ \cite{key-11,key-12,key-13,key-14,key-24}.
Then, the ratio $\frac{T_{E}(\omega)}{T_{H}}=\frac{2M}{2M-\omega}$
represents the deviation of the BH radiation spectrum from the strictly
thermal character \cite{key-11,key-12,key-13,key-14,key-24}. In correspondence
of $T_{E}(\omega)$ one can also introduce the \emph{effective mass
}and of the \emph{effective horizon} \cite{key-11,key-12,key-13,key-14,key-24}
\begin{equation}
M_{E}\equiv M-\frac{\omega}{2},\mbox{ }r_{E}\equiv2M_{E}\label{eq: effective quantities}
\end{equation}

\noindent of the BH \emph{during} the emission of the particle, i.e.
\emph{during} the BH contraction phase \cite{key-11,key-12,key-13,key-14,key-24}.
Such quantities are average values of the mass and the horizon \emph{before}
and \emph{after} the emission \cite{key-11,key-12,key-13,key-14,key-24}.

\noindent The correction to the thermal spectrum is also very important
for the physical interpretation of BH QNMs, and, in turn, is very
important to realize the underlying quantum gravity theory as BHs
represent theoretical laboratories for developing quantum gravity
and BH QNMs are the best candidates like quantum levels \cite{key-11,key-12,key-13,key-14,key-24}. 

In the appendix of \cite{key-13} we have rigorously shown that, if
one takes into account the deviation from the strictly thermal behavior
of the spectrum, eq. (\ref{eq: quasinormal modes}) must be replaced
with 
\begin{equation}
\begin{array}{c}
\omega_{n}=\ln3\times T_{E}(\omega_{n})+2\pi i(n+\frac{1}{2})\times T_{E}(\omega_{n})+\mathcal{O}(n^{-\frac{1}{2}})=\\
\\
=\frac{\ln3}{4\pi\left[2M-(\omega_{0})_{n}\right]}+\frac{2\pi i}{4\pi\left[2M-(\omega_{0})_{n}\right]}(n+\frac{1}{2})+\mathcal{O}(n^{-\frac{1}{2}}).
\end{array}\label{eq: quasinormal modes corrected}
\end{equation}
In other words, the Hawking temperature $T_{H}$ is replaced by the
effective temperature $T_{E}$ in eq. (\ref{eq: quasinormal modes}).
We can also give an intuitive but very enlightening physical interpretation
of such a replacing. Considering the deviation from the thermal spectrum
it is natural to assume that the Euclidean BH solution converges now
to a \emph{non-thermal} circle at infinity \cite{key-11,key-12}.
Therefore, it is straightforward the replacement \cite{key-11,key-12}

\noindent 
\begin{equation}
\beta_{H}=\frac{1}{T_{H}}\rightarrow\beta_{E}(\omega)=\frac{1}{T_{E}(\omega)},\label{eq: sostituiamo}
\end{equation}

\noindent which takes into account the deviation of the radiation
spectrum of a black hole from the strictly thermal feature. In this
way, the spacing of the poles in eq. (\ref{eq: quasinormal modes corrected})
coincides with the spacing \cite{key-11,key-12}

\noindent 
\begin{equation}
2\pi iT_{E}(\omega)=2\pi iT_{H}(\frac{2M}{2M-\omega}),\label{eq: spacing}
\end{equation}

\noindent expected for a \emph{non-thermal} Green's function (a dependence
on the frequency is present) \cite{key-11,key-12}.

Strictly speaking, eqs. (\ref{eq: quasinormal modes}) and (\ref{eq: quasinormal modes corrected})
are corrected only for scalar and gravitational perturbations. On
the other hand, for large $n$ eq. (\ref{eq: quasinormal modes corrected})
is well approximated by (we consider the leading term in the imaginary
part of the complex frequencies) 
\begin{equation}
\omega_{n}\simeq\frac{2\pi in}{4\pi\left[2M-(\omega_{0})_{n}\right]},\label{eq: andamento asintotico}
\end{equation}
and we have shown in the appendix of \cite{key-13} that the behavior
(\ref{eq: andamento asintotico}) also holds for $j=1$ (vector perturbations).
In complete agreement with Bohr's correspondence principle, it is
trivial to adapt the analysis in \cite{key-1} in the sense of the
Appendix of \cite{key-13} and, in turn, to show that the behavior
(\ref{eq: andamento asintotico}) holds if $j$ is a half-integer
too. In \cite{key-11,key-12,key-13} we have shown that the physical
solution of (\ref{eq: andamento asintotico}) for the absolute values
of $\omega_{n}$ is 
\begin{equation}
(\omega_{0})_{n}=M-\sqrt{M^{2}-\frac{n}{2}}.\label{eq: radice fisica}
\end{equation}
Now, we clarify how the correspondence between QNMs and Hawking radiation
works. One considers a BH original mass $M.$ After an high number
of emissions of Hawking quanta and eventual absorptions, because neighboring
particles can, in principle, be captured by the BH, the BH is at an
excited level $n-1$ and its mass is $M_{n-1}\equiv M-(\omega_{0})_{n-1}$
where $(\omega_{0})_{n-1},$ is the absolute value of the frequency
of the QNM associated to the excited level $n-1.\;$ $(\omega_{0})_{n-1}\;$
is interpreted as the total energy emitted at that time. The BH can
further emit a Hawking quantum to jump to the subsequent level: $\triangle M_{n}=(\omega_{0})_{n-1}-(\omega_{0})_{n}.$
Now, the BH is at an excited level $n$ and the BH mass is 
\begin{equation}
\begin{array}{c}
M_{n}\equiv M-(\omega_{0})_{n-1}+\triangle M_{n}=\\
\\
=M-(\omega_{0})_{n-1}+(\omega_{0})_{n-1}-(\omega_{0})_{n}=M-(\omega_{0})_{n}.
\end{array}\label{eq: masse}
\end{equation}
The BH can, in principle, return to the level $n-1$ by absorbing
an energy $-\triangle M_{n}=(\omega_{0})_{n}-(\omega_{0})_{n-1}$.
By using eq. (\ref{eq: radice fisica}) one gets immediately \cite{key-11,key-12,key-13}
\begin{equation}
\triangle M=\omega=(\omega_{0})_{n-1}-(\omega_{0})_{n}=-f_{n}(M,n)\label{eq: variazione}
\end{equation}
with \cite{key-11,key-12,key-13} 
\begin{equation}
f_{n}(M,n)\equiv\sqrt{M^{2}-\frac{1}{2}(n-1)}-\sqrt{M^{2}-\frac{n}{2}}.\label{eq: f(M,n)}
\end{equation}
One can easily check that in the very large $n$ limit one gets $f_{n}(M,n)\rightarrow\frac{1}{4M}$.
Thus, by using eq. (\ref{eq: variazione area}) one gets immediately
that in the very large $n$ limit two adjacent QNMs resolve an elementary
edge (or site) of a quantized surface $\triangle A\rightarrow8\pi$,
which corresponds to the famous historical result by Bekenstein \cite{key-15}
and this cannot be a coincidence. Then, the quantum levels are equally
spaced for both emissions and absorptions being $\frac{1}{4M}$ the
jump between two adjacent levels, and this also clearly falsifies
the criticism 1. by Khriplovich because in our semiclassical approximation
with $n\gg1$, the frequencies which corresponds to transitions between
energy levels with $\triangle n\ll n$ are integer multiples of the
classical frequency $\omega=\frac{1}{4M}$. 

The BH model that we re-analysed here is somewhat similar to the semi-classical
Bohr model of the structure of a hydrogen atom \cite{key-17,key-18}.
In our BH model during a quantum jump a discrete amount of energy
is indeed radiated and, for large values of the principal quantum
number $n,$ the analysis becomes independent of the other quantum
numbers. In a certain sense, QNMs represent the \textquotedbl{}electron\textquotedbl{}
which jumps from a level to another one and the absolute values of
the QNMs frequencies represent the energy \textquotedbl{}shells\textquotedbl{}.
In Bohr model \cite{key-17,key-18} electrons can only gain and lose
energy by jumping from one allowed energy shell to another, absorbing
or emitting radiation with an energy difference of the levels according
to the Planck relation $E=hf$, where $\: h\:$ is the Planck constant
and $f\:$ the transition frequency. In our BH model, QNMs can only
gain and lose energy by jumping from one allowed energy shell to another,
absorbing or emitting radiation (emitted radiation is given by Hawking
quanta) with an energy difference of the levels according to eq. (\ref{eq: variazione}).
The similarity is completed if one note that the interpretation of
eq. (\ref{eq: radice fisica}) is of a particle, the ``electron'',
quantized with anti-periodic boundary conditions on a circle of length
\cite{key-11,key-12,key-13} 
\begin{equation}
L=\frac{1}{T_{E}(E_{n})}=4\pi\left(M+\sqrt{M^{2}-\frac{n}{2}}\right),\label{eq: lunghezza cerchio}
\end{equation}
which is the analogous of the electron travelling in circular orbits
around the hydrogen nucleus, similar in structure to the solar system,
of Bohr model \cite{key-17,key-18}. Clearly, all these similarities
with the Bohr semi-classical model of the hydrogen atom and all these
consistences with well known results in the literature of BHs, starting
by the universal Bekenstein's result, \emph{cannot} be coincidences,
but are confirmations of the correctness of the current analysis.

It is also important to stress that the Bohr-like BH has important
implications for the BH information paradox \cite{key-25}. In fact,
the BH seems to be a well defined quantum mechanical system, having
an ordered, discrete quantum spectrum. This is surely consistent with
the unitarity of the underlying quantum gravity theory and with the
idea that information should come out in BH evaporation. We have indeed
recently shown that the time evolution of the Bohr-like BH obeys a
\emph{time dependent Schrödinger equation }\cite{key-26}. In that
way, the physical state and the correspondent\emph{ wave function
}are written in terms of an \emph{unitary} evolution matrix instead
of a density matrix \cite{key-26}. Thus, the final state results
to be a \emph{pure} quantum state instead of a mixed one while the
entanglement problem connected with the information paradox is solved
showing that the emitted radiation is entangled with BH QNMs \cite{key-26}.

\section{Conclusion remarks}

Khriplovich \cite{key-20} verbatim claimed that ``the correspondence
principle does not dictate any relation between the asymptotics of
quasinormal modes and the spectrum of quantized black holes'' and
that ``this belief is in conflict with simple physical arguments''.
In this paper we have shown that the criticisms in \cite{key-20}
work only for the original proposal by Hod, while they do not work
for the improvements suggested by Maggiore and recently finalized
by the author and collaborators through a connection between Hawking
radiation and BH QNMs. Thus, QNMs can be really interpreted as BH
quantum levels in a Bohr-like model for BHs where QNMs represent the
\textquotedbl{}electron\textquotedbl{} which jumps from a level to
another one and the absolute values of the QNMs frequencies represent
the energy \textquotedbl{}shells\textquotedbl{}. Then, the emission
of Hawking quanta and the absorptions of particles represent the jumps
among the various quantum levels. 

The results in this paper have also important consequences on the
BH information paradox.

\section{Acknowledgements }

The Scuola Superiore Internazionale di Studi Universitari e Ricerca
``Santa Rita'' has to be thanked for supporting this paper. I thank
two unknown referees for useful comments.

\end{document}